\documentclass[12pt,a4paper]{article}
\usepackage{amsmath}
\begin{document}

\title{\Large The Central Error of  M{\small YRON} W. E{\small VANS}' ECE Theory -\\
a Type Mismatch}

\author{Gerhard W. Bruhn\\
Darmstadt University of Technology, D 64289 Darmstadt\\
bruhn@mathematik.tu-darmstadt.de}

\maketitle

\begin{abstract} 
In Section 1 we give a sketch of the basics of spacetime manifolds. Namely the tetrad coefficients
$q_{\bf a}^\mu,$ 
are introduced which {\sc M.W. Evans}
believes to be an essential tool of argumentation leading far beyond 
the limitations of General Relativity
because of giving the opportunity of modelling several other force fields of
modern physics in addition to gravitation. However, as we shall see in Section 2,
the main errors of that ``theory'' are {\it invalid} field definitions: 
They are simply invalid and therefore useless due to {\it type mismatch}. 
This is caused by {\sc M.W. Evans}' {\bf bad habit} of suppressing 
seemingly unimportant indices.
{\bf There is no possibility of removing the tetrad indices a,b from  
M.W. E{\scriptsize VANS}'
field theory, i.e. the ECE Theory cannot be repaired.}
In Section 3 {\sc M.W. Evans}' concept of a non-Minkowskian spacetime manifold 
[1; Sect.2],[2; Chap.3.2], is shown to be erroneous. 
In Section 4 another erroneous claim of [1; Sect.3],
[2; Chap.3.3] is discussed.
\end{abstract}

\vspace*{2cm}

The following review of {\sc M.W. Evans}' 
{\large \bf E}instein 
{\large \bf C}artan 
{\large \bf E}vans field theory
refers to {\sc M.W. Evans}' FoPL article [1]. About one year later he took
the article over into his book [2] without essential changes.
Below the labels of type $(3.\cdot )/(\cdot)$ refer to [2]/[1] respectively.

%%%%%%%%%%%%%%%%%%%%%%%%%%%%%%%%%%%%%%%%%%%%%%%%%%%%%%%%%

\section{What M.W. E{\small VANS} should have given first: 
A clear description of his basic assumptions}

{\sc M.W. Evans} constructs his spacetime by a dubious {\it alternative} 
method to be discussed in Section 3 . 
Here we sketch the {\it usual} method of constructing the 4-dimensional 
spacetime manifold $\cal M$. 
The tangent spaces ${\bf T}_P$ at the points $P$ of $\cal M$ 
are spanned by the tangential basis vectors 
${\bf e}_\mu = \partial_\mu ~(\mu=0,1,2,3)$
at the respective points $P$ of $\cal M$.\\

There is a pseudo-metric defined at the points $P$ of $\cal M$ 
as a bilinear function 
$g : {\bf T}_P \times  {\bf T}_P \rightarrow  \bf R$. 
Therefore we can define the  matrix 
$(g_{\mu\nu})$ by 

\begin{equation}\label{(1.1)}
g_{\mu\nu} := g({\bf e}_\mu,{\bf e}_\nu) ,\end{equation}

which is assumed to be of Lorentzian signature, i.e.
there exist vectors 
$\bf e_a ~(a = 0, 1, 2, 3)$
in each ${\bf T}_P$ such that we have
$g(\bf{e_a,e_b}) = \eta_{\bf ab}$
where the matrix 
$(\eta_{\bf ab})$ 
is the Minkowskian diagonal matrix 
$diag(-1, +1, +1, +1)$. 
We say also the signature of the metric 
$(g_{\mu\nu})$ 
is supposed to be Lorentzian, i.e.
$(-,+,+,+)$.\\

A linear transform 
$L: {\bf T}_P \rightarrow  {\bf T}_P$
that fulfils 
$g(L{\bf e_a},L{\bf e_b}) = g({\bf e_a},{\bf e_b})$ 
is called a (local) Lorentz transform.
The Lorentz transforms of 
${\bf T}_P$ 
constitute the well-known
(local) Lorentz group. All Lorentz-transforms have the property 
$g(L{\bf V},L{\bf W}) = g({\bf V, W})$
for arbitrary vectors 
${\bf V, W}$ in ${\bf T}_P$.\\

Each set of orthonormalized vectors 
${\bf e_a~(a = 0, 1, 2, 3)}$,
in ${\bf T}_P$ is called a {\it tetrad} at the point $P$. We assume that 
a certain tetrad being chosen at each 
${\bf T}_P$ 
of the manifold $\cal M$. 
Then we have linear representations of the coordinate basis vectors 
${\bf e}_\mu = \partial_\mu\ (\mu=0,1,2,3)$ 
by the tetrad vectors at $P$:

\begin{equation}\label{(1.2)}
{\bf e}_\mu 
= q_\mu^{\bf a}\ {\bf e_a}.
\end{equation}

 From (\ref{(1.1)}) and (\ref{(1.2)}) we obtain due to the bilinearity of 
$g(\cdot,\cdot)$
 
\begin{equation}\label{(1.3)}
g_{\mu\nu} = g({\bf e}_\mu,{\bf e}_\nu) = q_\mu^{\bf a}\ q_\nu^{\bf b}\
g({\bf e_a, e_b}) 
= q_\mu^{\bf a}\ q_\nu^{\bf b}\ \eta_{\bf ab}.
\end{equation}

The matrix  
$(g_{\mu\nu})$ 
is symmetric therefore. And more generally also 
$g({\bf V, W}) = g({\bf W,V})$ 
for arbitrary vectors 
${\bf V, W}$
of ${\bf T}_P$.
In addition, the multiplication theorem for determinants yields the matrix 
$(g_{\mu\nu})$ 
to be nonsingular.\\
 
A (non-Riemannian) linear connection is supposed, i.e. we have 
{\it covariant} derivatives 
$D_\mu$  
in direction of 
${\bf e}_\mu$ 
given by
 
\begin{equation}\label{(1.4)}
D_\mu F := \partial_\mu F 
\end{equation}

for functions 
$F$ 
($=(0,0)$-tensors), while a $(1,0)$-tensor 
$F^\nu$ 
has the derivative
 
\begin{equation}\label{(1.5)}
D_\mu F^\nu
:= \partial_\mu F^\nu
+
\Gamma_{\mu\ \rho}^{\ \nu}\ F^\rho 
\end{equation}

and for a $(0,1)$-tensor 
$F_\nu$ 
we have

\begin{equation}\label{(1.6)}
D_\mu F_\nu 
:= \partial_\mu F_\nu 
-
\Gamma_{\mu\ \nu}^{\ \rho}\ F_\rho.
\end{equation}

For coordinate dependent quantities the connection causes the additional 
terms in Eqns.(\ref{(1.5)}-\ref{(1.6)}) with the coefficients 
$\Gamma_{\mu\ \nu}^{\ \rho}$.\\

By the analogue way the connection gives rise to additional terms
with coefficients 
$\omega_{\mu\ \bf b}^{\ \bf a}$
for the covariant derivatives of tetrad dependent quantities, namely   
 
\begin{equation}\label{(1.7)}
D_\mu F^{\bf a} := \partial_\mu F^{\bf a}
+
\omega_{\mu\ \bf b}^{\ \bf a}\ F^{\bf b} 
\end{equation}

and 
\begin{equation}\label{(1.8)}
D_\mu F_{\bf a}
:= \partial_\mu F_{\bf a}
-
\omega_{\mu\ \bf a}^{\ \bf b}\ 
F_{\bf b} .
\end{equation}

%%%%%%%%%%%%%%%%%%%%%%%%%%%%%%%%%%%%%%%%%%%%%%%%%%

\section{M.W. E{\small VANS}' Generally Covariant Field Equation}

{\sc M.W. Evans} starts with Einstein's Field equation
 
\begin{equation}\label{(2.1)}
R^{\mu\nu} - \frac{1}{2} R\ g^{\mu\nu}
= T^{\mu\nu}
\end{equation}
which is ``multiplied'' by 
$q_\nu^{\bf b}\ \eta_{\bf ab}$ 
to obtain

\begin{equation}
R_{\bf a}^\mu - \frac{1}{2} R\ q_{\bf a}^\mu
= T_{\bf a}^\mu \label{(2.2)}.
\end{equation}

Here he suppresses the tetrad index 
${\bf a}$:\\

Quote from [2]/[1]\\

$$(3.18)/(16)\hspace{4cm}
R^\mu - \frac{1}{2} R\ q^\mu
= T^\mu \hspace*{30cm}
$$
%}

He now ``wedges'' that by 
$q_{\bf b}^\nu$ 
to obtain

\begin{equation}\label{(2.4)}
R_{\bf a}^\mu 
\wedge 
q_{\bf b}^\nu
- 
\frac{1}{2} R\ q_{\bf a}^\mu 
\wedge 
q_{\bf b}^\nu
= T_{\bf a}^\mu
\wedge  
q_{\bf b}^\nu .
\end{equation}

Here he suppresses the tetrad indices 
${\bf a},{\bf b}$ again:\\

Quote from [2]/[1]:

$$
(3.25)/(23)\hspace{3cm}
R^\mu \wedge q^\nu
- 
\frac{1}{2} R q^\mu \wedge q^\nu
= T^\mu \wedge  q^\nu \hspace{6cm}
$$\\

{\bf Remark}\\

The wedge product used by  {\sc M.W. Evans} here is the wedge product of vectors
${\bf A} = A^\mu {\bf e}_\mu$:

$$
{\bf A \wedge B}
= \frac{1}{2}
(A^\mu B^\nu - A^\nu B^\mu)\ {\bf e}_\mu \wedge \bf e_\nu
$$

written in short hand as
$$\hspace*{4cm}
A^\mu \wedge B^\nu
:=
\frac{1}{2}
(A^\mu B^\nu - A^\nu B^\mu) .\hspace{4cm}\rule{2mm}{2mm}
$$\\

{\sc M.W. Evans} remarks the term 
$R^\mu \wedge q^\nu$
to be antisymmetric like the electromagnetic field tensor 
$G^{\mu\nu}.$ 
Hence he feels encouraged to try the following ansatz\\

Quote from [2]/[1]:
%\hline
$$(3.29)/(27)\hspace{2cm}
G^{\mu\nu} 
= 
G^{(0)} (R^{\mu\nu(A)} - 
\frac{1}{2} R\ q^{\mu\nu(A)})\hspace*{6cm}
$$
%}

where

$$(3.26-27)/(24-25)\hspace{1cm}
R^{\mu\nu(A)} = R^\mu \wedge q^\nu,\hspace{1cm}
q^{\mu\nu(A)} = q^\mu \wedge q^\nu .\hspace*{6cm}
$$

Thus, {\sc M.W. Evans}' ansatz (3.29)/(27) {\it with written tetrad indices} is
 
\begin{equation}\label{(2.5)}
G^{\mu\nu} 
=
G^{(0)} (R_{\bf a}^\mu \wedge q_{\bf b}^\nu
- 
\frac{1}{2} R\ q_{\bf a}^\mu \wedge q_{\bf b}^\nu).
\end{equation}

However, by comparing the left hand side and the right hand side 
it is evident that the ansatz cannot be correct due to {\it type mismatch}:
The tetrad indices $\bf a$ and $\bf b$ are not available at the left hand 
side, which means that both sides have different transformation properties.\\

%\fbox{\fbox{
{\large{\bf M.W. E{\small VANS}' field ansatz (3.29)/(27) 
is unjustified due to type mismatch.}}\\
%}}

The tetrad indices 
$\bf a,\bf b$
must be removed {\it legally}. 
The only way to do so is to sum over 
$\bf a,\bf b$
with some weight factors 
$\chi^{\bf ab}$, i.e. to insert a factor
$\chi^{\bf ab}$ 
on the right hand side of (3.29)/(27), at (\ref{(2.5)}) 
in our detailed representation.
Our first choice for 
$\chi^{\bf ab}$ 
is the Minkowskian 
$\eta^{\bf ab}$. 
However, then the right hand side of (3.29)/(27) vanishes since we have

\begin{equation}\label{(2.6)}
q_{\bf a}^\mu \wedge q_{\bf b}^\nu\ \eta^{\bf ab} 
=
q_{\bf a}^\mu\ q_{\bf b}^\nu\ \eta^{\bf ab} - q_{\bf a}^\nu\ q_{\bf b}^\mu\ \eta^{\bf ab}
= g^{\mu\nu} - g^{\nu\mu} = 0
\end{equation}

and

\begin{equation}\label{(2.7)}
R_{\bf a}^\mu \wedge q_{\bf b}^\nu\ \eta^{\bf ab} 
=
R_{\bf a}^\mu\ q_{\bf b}^\nu\ \eta^{\bf ab}
-
R_{\bf a}^\nu\ q_{\bf b}^\mu\ \eta^{\bf ab}
=
R^{\mu\nu} - R^{\nu\mu} = 0 
\end{equation}

due to the symmetry of the metric tensor 
$g^{\mu\nu}$
and of the Ricci tensor 
$R^{\mu\nu}$ 
[4; (3.91)].\\
 
One could try to find a matrix 
$(\chi^{\bf ab})$ 
different from the Minkowskian to remove the indices ${\bf a, b}$ from 
 equations (3.25-29)/(23-27).  
That matrix should not depend on the special 
tetrad under consideration i.e. be invariant under arbitrary
Lorentz transforms $L$:
 
\begin{equation}\label{(2.8)}
L_{\bf c}^{\bf a}\ 
\chi^{\bf cd}\ 
L{\bf_d^b} 
= \chi^{\bf ab}
 \qquad \mbox{where}\quad 
L\ {\bf e}_{\bf a} =: L_{\bf a}^{\bf b}\ {\bf e_b}.
\end{equation}

However, due to the definition of the Lorentz transforms the matrices
$\lambda\ (\eta^{\bf ab})$ 
with some factor $\lambda$  are the only 
matrices with that property.\\

Therefore we may conclude that only a trivial {\bf zero} em-field 
$G^{\mu\nu}$ 
can fulfil the {\it corrected} {\sc M.W. Evans} field ansatz. \\

%\fbox
{\large\bf{The correction of M.W. E{\small VANS}' antisymmetric field ansatz 
(3.29)/(27) 
 yields the trivial zero case merely and is irreparably therefore.}}

%%%%%%%%%%%%%%%%%%%%%%%%%%%%%%%%%%%%%%%%%%%%%%%%%%%%%%%%%%%%%%%

\section{Further Remarks}

The following remarks are concerning {\sc M.W. Evans}' idea of the spacetime manifold as 
represented in his [2; Chap.3.2]/[1; Sec.2].\\

He starts with a curvilinear parameter representation 
${\bf r = r}(u_1, u_2, u_3)$
in a space the property of which is not explicitely described but turns 
out to be an Euclidean ${\bf R}^3$ due to the Eqns.(3.10)/(8) below. \\

Quote from [2]/[1]:\\

{\it 
Restrict attention initially to three non-Euclidean space dimensions. 
The set of curvilinear coordinates is defined as 
$(u_1, u_2, u_3)$, 
where the functions are single valued and continuously differentiable, 
and where there is a one to one relation between 
$(u_1, u_2, u_3)$ 
and the Cartesian coordinates. The position vector is 
${\bf r}(u_1, u_2, u_3)$, 
and the arc length is the modulus of the infinitesimal displacement 
vector:

$$(3.7)/(5)\qquad\qquad
ds = \left| d{\bf r} \right| = \left| \frac{\partial{\bf r}}{\partial u_1} du_1 
+      \frac{\partial{\bf r}}{\partial u_2} du_2 
+      \frac{\partial{\bf r}}{\partial u_3} du_3 \right|.\hspace*{6cm} $$

The metric coefficients are 
$\frac{\partial{\bf r}}{\partial u_i}$, 
and the scale factors are:

$$(3.8)/(6)\hspace{4cm}
h_i = \left| \frac{\partial{\bf r}}{\partial u_i} \right| .\hspace*{6cm} 
$$

The unit vectors are
$$(3.9)/(7)\hspace{4cm}
{\bf e}_i = \frac{1}{h_i}
\frac{\partial{\bf r}}{\partial u_i}\hspace*{6cm}
$$

and form the 
$O(3)$ 
symmetry cyclic relations:

$$(3.10)/(8)\qquad\qquad
{\bf e_1 \times e_2 = e_3},\hspace{0.5cm} 
{\bf e_2 \times e_3 = e_1},\hspace{0.5cm} 
{\bf e_3 \times e_1 = e_2},\hspace*{6cm} 
$$

where 
$O(3)$ 
is the rotation group of three dimensional space [3-8]. 
The curvilinear coordinates are orthogonal if:

$$(3.11)/(9)\qquad\qquad\quad
{\bf e_1 \cdot e_2} = 0,\hspace{0.5cm} 
{\bf e_2 \cdot e_3} = 0,\hspace{0.5cm} 
{\bf e_3 \cdot e_1} = 0.\hspace*{6cm} 
$$

The symmetric metric tensor is then defined through the line element, 
a one form of differential geometry\\
 
\hspace*{3cm}\fbox{{\large \bf NO! A symmetric TWO-form}}:

$$(3.12)/(10)\hspace{4cm}
\omega_1 = ds^2  = q^{ij(S)} du_idu_j ,\hspace*{6cm} 
$$

and the anti-symmetric metric tensor through the area element, 
a two form of differential geometry:

$$(3.13)/(11)\hspace{4cm}
\omega_2 = dA = - \frac{1}{2} q^{ij(A)} du_i \wedge du_j .\hspace*{6cm}
$$

These results generalize as follows to the four dimensions of any 
non-Euclidean space-time:

$$(3.14)/(12)\hspace{4cm}
\omega_1 = ds^2 = q^{\mu\nu(S)} du_\mu du_\nu ,\hspace*{8cm} 
$$

$$(3.15)/(13)\hspace{3cm}
\omega_2 =\ ^\star \omega_1 = dA = - \frac{1}{2} q^{\mu\nu(A)} 
du_\mu \wedge du_\nu.\hspace*{8cm}$$\\
\hspace*{4cm}\fbox{\large{\bf{WRONG HODGE DUALITY!}}}.\\

In differential geometry the element $du_\sigma$ 
is dual to the wedge product $du_\mu\wedge\nolinebreak[4]du_\nu$.\\
\hspace*{4cm}\fbox{\large{\bf{WRONG! NOT in 4-D}}}.\\ 

The symmetric metric tensor is:

\[(3.16)/(14)\hspace{2cm}
q^{\mu\nu(S)}
= \left[\begin{array}{cccc}
 h_0^2  & h_0 h_1 & h_0 h_2 & h_0 h_3 \\
h_1 h_0 &  h_1^2  & h_1 h_2 & h_1 h_3 \\
h_2 h_0 & h_2 h_1 &  h_2^2  & h_2 h_3 \\
h_3 h_0 & h_3 h_1 & h_3 h_2 &  h_3^2
\end{array} \right]\hspace*{8cm}
\]

and the anti-symmetric metric tensor is:

\[(3.17)/(15)\hspace{2cm}
q^{\mu\nu(A)}
= \left[\begin{array}{cccc}

   0    & -h_0 h_1 & -h_0 h_2 & -h_0 h_3 \\
h_1 h_0 &     0    & -h_1 h_2 &  h_1 h_3 \\
h_2 h_0 &  h_2 h_1 &     0    & -h_2 h_3 \\
h_3 h_0 & -h_3 h_1 &  h_3 h_2 &     0
\end{array} \right]\hspace*{8cm}
\]}

\nopagebreak[4]
(End of quote)\\ 

The symmetric metric (3.16)/(14) cannot be correct since 
having a {\it vanishing} determinant: All line vectors are parallel. 
The reason is that the author {\sc M.W. Evans} has
forgotten to insert the scalar products of his basis vectors.
A similiar argument holds for Equ.(3.17)/(15) being dubious.\\

However, even if one avoids all possibilities mentioned above of 
going astray
{\sc M.W. Evans}' method has {\it one crucial shortcoming}: The metric definable by 
that method. As follows from (3.7)/(5) we have 
$ds^2 \ge 0$, 
i.e. the metric is {\it positive definite}. That is a heritage of {\sc M.W. Evans}' 
construction of spacetime as an embedding into a real Euclidian  space 
(defining the metric by (3.7)/(5)) that one cannot get rid off.\\

{\large{\bf M.W. E{\small VANS}' construction cannot yield a spacetime with
local Minkowskian i.e. {\it indefinite} metric.}}\\

That was the reason why we sketched a  correct method of constructing the
spacetime manifold of General Relativity at the beginning of this article in Section 1. 
{\sc M.W. Evans}' alternative method of [2; Chap.3.2]/[1; Chap.2] is useless.

%---------------------------------------------%

\section{A Remark on [2; Chap.3.4]/[1; Sect.4]}

With\\

Quote from [2]/[1]\\
$$
(3.2)/(43)\hspace{4cm}
R^\mu = \alpha\ q^\mu \hspace*{5cm}\hspace{8cm}
$$
%}

 claims {\it proportionality} between the tensors 
$R_{\bf a}^\mu$ 
and 
$q_{\bf a}^\mu$:

\begin{equation}\label{(4.1)}
R_{\bf a}^\mu = \alpha\ q_{\bf a}^\mu .
\end{equation}

However, there is no proof in [2; Chap.3.4]/[1; Sect.4] available. 
Indeed, if we assume eq.(\ref{(4.1)}) then we obtain the curvature

\begin{equation}\label{(4.2)}
R = R^{\mu\nu}\ g_{\mu\nu} 
= (R_{\bf a}^\mu\ \eta^{\bf ab}\ q_{\bf b}^\nu)\ g_{\mu\nu}
=
R_{\bf a}^\mu\ q_\mu^{\bf a}\\
= \alpha\ q_{\bf a}^\mu\ q_\mu^{\bf a} = 4\ \alpha ,
\end{equation}

but the equation
$R_{\bf a}^\mu\ q_\mu^{\bf a} = \alpha\ q_{\bf a}^\mu\ q_\mu^{\bf a}$
may have other solutions than eq.(\ref{(4.1)}). 
Hence there is no way from eq.(\ref{(4.2)}) back to eq.(\ref{(4.1)}).\\

%\fbox
{\large{\bf The considerations of [2; Chap.3.4]/[1; Sect.4] may be 
based on a logical flaw.}}

%---------------------------------------------%

\end{document}